\documentclass[]{JHEP3}
\usepackage{epsfig}
\def\be{\begin{equation}}
\def\ee{\end{equation}}
\def\bea{\begin{array}}
\def\eea{\end{array}}
\def\beqa{\begin{eqnarray}}
\def\eeqa{\end{eqnarray}}
\def\beqas{\begin{eqnarray*}}
\def\eeqas{\end{eqnarray*}}

\def\bp{\begin{picture}}
\def\ep{\end{picture}}
\def\bc{\begin{center}}
\def\ec{\end{center}}
\def\bfig{\begin{figure}}
\def\efig{\end{figure}}

\def\bit{\begin{itemize}}
\def\eit{\end{itemize}}
\def\nn{\nonumber}
\def\f{\frac}

\def\[{\left[}
\def\]{\right]}
\def\({\left(}
\def\){\right)}

\def\..{\left.}
\def\.{\right.}
\def\tl{\tilde}
\def\ra{\rightarrow}
\def\la{\leftarrow}

\def\tm{\times}

\def\da{\dagger}

\def\la{\lambda}

\def\al{\alpha}

\def\ka{\kappa}

\def\ep{\epsilon}

\def\pa{\partial}
\def\pr{\prime}

\title{ ExtraOrdinary Gauge Mediation Extension of deflected AMSB}
\author{Fei Wang$^1$, \\
$^1$ School of Physics, Zhengzhou University, 450000,ZhengZhou
P.R.China
}

\abstract{ Extraordinary gauge mediation extension of deflected AMSB scenarios can be interesting because it can accommodate together the deflection in the Kahler potential and the superpotential. We revisit the EGM scenario and derive the analytical expressions for soft SUSY breaking parameters in EGM and EGM extension of deflected AMSB scenarios with wavefunction renormalization approach, especially the case with vanishing gauge beta-function at an intermediate energy scale. The Landau pole and proton decay constraints are also discussed.}

\begin{document}
\maketitle \indent
\newpage
\section{Introduction}
Weak scale supersymmetry(SUSY), which is a leading candidate for physics beyond the standard model(SM), can solve elegantly the hierarchy problem of the Higgs boson by introducing various superpartners at TeV scale. Besides, gauge couplings unification, which can not be exact in SM, can be successful in its SUSY extensions. The dark matter(DM) puzzle as well as the BAU(baryon-asymmetric universe) puzzle etc, can also be explained with proper DM candidates and baryogensis mechanisms in SUSY. It is worth to note that the Higgs scalar, which was discovered by the ATALS\cite{ATLAS:higgs} and CMS\cite{CMS:higgs} collaborations of LHC in 2012, lie miraculously in the small $'115-135'$ GeV window predicted by the low energy SUSY. 


It is well known that the soft SUSY spectrum, including the gaugino and sfermion masses, are determined by the SUSY breaking mechanism.
Depending on the way  the visible sector $'feels'$ the SUSY breaking effects in the hidden sector, the SUSY breaking mechanisms can be classified into gravity mediation\cite{SUGRA}, gauge mediation\cite{GMSB}, anomaly mediation\cite{AMSB} scenarios, etc. Gauge mediated SUSY breaking(GMSB) scenarios, which will not cause flavor and CP problems that bothers gravity mediation models, are calculable, predictive, and phenomenologically distinctive with minimal messenger sector. However, it is difficult for minimal GMSB to explain 125 GeV Higgs with TeV scale soft SUSY breaking parameters because of the vanishing trilinear terms at the messenger scale. An interesting extension is the (extra)ordinary gauge mediation (EGM) scenarios\cite{EGM,Dumitrescu:2010ha}, in which the messenger sector can include all renormalizable, gauge invariant couplings between the messengers and any number of singlet fields. In fact, many examples in the literature of OGM deformed by mass terms can fall into this category and their generic properties can be obtained therein.

  Gravity, which can couple to everything, can generate the soft masses by the auxiliary field of the gravitational multiplet.
Such a $'pure'$ gravity mediation scenario with negligible contributions from direct non-renormalizable contact terms is called the anomaly mediated SUSY breaking(AMSB). Pure anomaly mediation is bothered by the tachyonic slepton problem \cite{tachyonslepton}. Its non-trivial extensions with messenger sectors, namely the deflected AMSB\cite{dAMSB,okada,fei}, can elegantly solve such tachyonic slepton problems through the deflection of the renormalization group equation (RGE) trajectory \cite{deflect:RGE-invariance}. There are two types of deflections in the literature, namely, the deflection in the superpotential or deflection in the Kahler potential\cite{Nelson:2002sa,luty,luty2,fei:kahlerdeflection}. However, it is difficult to determined consistently the deflection parameter $'d'$ and soft SUSY parameters if both deflections are present. We find that such a scenario can be seen as a special case of the EGM extension of deflected AMSB. Besides, the most generic extension of the messenger sector can also be interesting theoretically.

We revisit the EGM scenario and derive the analytical results for the GMSB contributions in EGM and EGM extension of deflected AMSB, especially the case with a vanishing beta function at an intermediate scale and the case with $rank(\la_{ij})=rank (m_{ij})=0$. Our result is especially useful for cases with hierarchical messenger scales. The extension of the EGM to deflected AMSB is straightforward. We show that such an extension can indeed accommodate both types of deflections in AMSB.

This paper is organized as follows. In Sec \ref{dAMSB}, we discuss the deflected AMSB scenario with EGM extension. In Sec \ref{EGM}, we discuss the analytical expressions of soft SUSY parameters in EGM. In Sec \ref{Landau}, constraints from the Landau pole  with multiple messengers are given. Sec \ref{conclusion} contains our conclusions.
\section{\label{dAMSB} Extraordinary gauge mediation in deflected AMSB}
To fully understand the deflected AMSB scenarios with the presence of both the superpotential and Kahler potential deflection, we need the deflection parameter $'d'$ to derive the soft SUSY breaking parameters. One can minimize the whole scalar potential to obtain $'d'$, but such a case by case study seems tedious. We find that the solution can be obtained in EGM extension of deflected AMSB.

In deflected AMSB, the Kahler potential can have the following types of deflection with holomorphic terms for messengers
\beqa
K\supseteq \ka_{ij}\f{\phi^\da}{\phi} \tl{P}_iP_j~,
\eeqa
or the deflection from the couplings in the superpotential
\beqa
W\supseteq \la_{ij} \tl{X} \tl{Q}_i Q_j +W(\tl{X})~,
\eeqa
with a proper form of superpotential $W(\tl{X})$ for the pseudo-moduli field $\tl{X}$ to determine the deflection parameter $\tl{d}$
\beqa
\f{F_{\tl{X}}}{\tl{X}}=(\tl{d}+1)F_\phi~.
\eeqa
We should note that the messengers $\tl{P}_i,P_i$ can possibly be identified to be the $\tl{Q}_i, Q_i$ superfields.
Combining both Kahler and superpotential deflection, we have
\beqa
{\cal L}\supseteq \int d^4\theta  \[\ka_{ij}\f{\phi^\da}{\phi} \tl{P}_iP_j\] +\int d^2\theta  \[\la_{ij}\tl{X} \tl{Q}_i Q_j +W(\tl{X})\]~,
\eeqa
which can be rewritten as
\beqa
{\cal L}\supseteq \int d^2\theta (\la_{ij} \tl{X} + \ka_{ij} \tl{T})\tl{\phi}_i \phi_j+W(\tl{X})~,
\eeqa
with $\tl{T}$ the auxiliary field with VEV
\beqa
\langle\tl{T}\rangle=F_\phi-\theta^2 F^2_\phi~,
  \eeqa
 and $\phi$ the conformal compensator field which carries the SUSY breaking information in the SUSY breaking sector
 \beqa
 \phi=1+\theta^2 F_\phi~.
 \eeqa

We can rotate $\tl{X}$ and $\tl{T}$ so that only one combination $X$ will acquire F-term VEVs while $T$ will acquire only the lowest component VEVs
\beqa
\label{XT}
X&=&\f{1}{\sqrt{F_{\tl{X}}^2+F_\phi^2}}\[ F_{\tl{X}} \tl{X}-F^2_\phi\tl{T}\]~,\nn\\
T&=&\f{1}{\sqrt{F_{\tl{X}}^2+F_\phi^2}} \[F_\phi^2 \tl{X} + F_{\tl{X}} \tl{T}\]~,
\eeqa
So the superpotential can rewritten as
\beqa
W\supseteq (\la_{ij} {X} + m_{ij})\tl{\phi}_i \phi_j~,
\eeqa
with $m_{ij}=\ka_{ij}\langle {T}\rangle$.

We will impose an non-trivial R-symmetry
\beqa
\left\{\bea{cc}&\la_{ij}\neq 0~,~~~~{\rm only~if}~~ R(\tl{\phi}_i)+  R({\phi}_j)=2-R(X)~, \\
&\ka_{ij}\neq 0~,~~~~{\rm only~if}~~ R(\tl{\phi}_i)+  R({\phi}_j)=2~,~~~~~~~~~~  \\
\eea\right.
\eeqa
as well as a symmetry for messengers to prevent destructive D-term contributions to sfermion masses.
After integrating out the messenger fields, the messenger determinant is proven by \cite{EGM} to be a monomial in $X$
\beqa
\label{det1}
\det \(\la_{ij}X+\ka_{ij}T\)=X^{n_0} G(\la, \ka T)~,
\eeqa
with
\beqa
n_0=\f{1}{R(X)}\sum\limits_{i=1}^n \[2-R(\tl{\phi}_i)+  R({\phi}_i)\].
\eeqa
Note that the most general deflection in AMSB is quite similar to that of EGM in GMSB except that the $X$ is given by
\beqa
{\sqrt{F_X^2+F_\phi^2}}\langle X \rangle= F_\phi\[(\tl{d}+1)\tl{X}^2-F_\phi^2\]+(F_{\tl{X}}^2+F_\phi^4)\theta^2,
\eeqa
from eqn.(\ref{XT}). Thus the effective deflection parameter $d$ is given by
\beqa
d&=&\f{(F_{\tl{X}}^2+F_\phi^4)}{F_\phi^2\[(\tl{d}+1)\tl{X}^2-F_\phi^2\]}-1,\nn\\
&=&\f{\[(\tl{d}+1)^2\tl{X}^2+F_\phi^2\]}{\[(\tl{d}+1)\tl{X}^2-F_\phi^2\]}-1,
\eeqa
For later convenience, the VEV of $X$ is denoted by $(X)=M+\theta^2 F_X$.

The soft SUSY breaking parameters in EGM extension of deflected AMSB can be obtained by wavefunction renormalization\cite{GMSB:wavefunction} approach
\bit
\item The gaugino masses are given as
\beqa
M_{i}&=& g_i^2\(\f{F_\phi}{2}\f{\pa}{\pa \ln\mu}-\f{d F_\phi}{2}\f{\pa}{\pa \ln |X|}\)\f{1}{g_i^2}(\mu,|X|)~,\nn\\
&=&g_i^2\(\f{F_\phi}{2}\f{\pa}{\pa \ln\mu}-\sum\limits_{a=1}^N \f{d F_\phi}{2}\f{\pa \ln M_a}{\pa \ln |X|}\f{\pa}{\pa \ln M_a}\)\f{1}{g_i^2}(\mu,M_a)~,
\eeqa

\item The trilinear couplings are given as
\beqa
A_0^{ijk}\equiv \f{A_{ijk}}{y_{ijk}}&=&\sum\limits_{i}\(-\f{F_\phi}{2}\f{\pa}{\pa\ln\mu}+\f{d F_\phi}{2}\f{\pa}{\pa\ln |X|}\) \ln \[Z_i(\mu,M_a)\]~.
\eeqa
So we need to calculate
\beqa
\f{\pa \ln Z_l(\mu,M_a)}{\pa \ln |X|} =\sum\limits_{M_j}\sum\limits_{g_i(\mu^\pr)}\f{\pa\ln M_j}{\pa \ln |X|}\f{\pa\ln g_i(\mu^\pr)}{\pa \ln M_j} \f{\pa \ln Z_l}{\pa\ln g_i(\mu^\pr)}+\[g_i(\mu^\pr)\ra y_a(\mu^\pr)\].
\eeqa
in which the sum over $g_i(\mu^\pr)$, which depends on the messenger thresholds $M_i$, take the values $g_i(M_1),g_i(M_2),\cdots,g_i(M_N),g_i(\mu)$.
The second term always vanishes because the anomalous dimension is continuous across the messenger thresholds.

\item The soft SUSY breaking scalar masses are given as
\beqa
m^2_{soft}&=&-\left|-\f{F_\phi}{2}\f{\pa}{\pa\ln\mu}+\f{d F_\phi}{2}\f{\pa}{\pa\ln |X|}\right|^2 \ln \[Z_i(\mu,M_a)\]~,\\
&=&-\(\f{F_\phi^2}{4}\f{\pa^2}{\pa (\ln\mu)^2}+\f{d^2F^2_\phi}{4}\f{\pa^2}{\pa(\ln |X|)^2}
-\f{d F^2_\phi}{2}\f{\pa^2}{\pa\ln|X|\pa\ln\mu}\) \ln \[Z_i(\mu,M_a)\],\nn
\eeqa
We need to calculate
\beqa
\f{\pa^2}{\pa \ln |X|^2}\ln Z_i(\mu,M_a)
&=&\[\f{\pa \ln M_j}{\pa \ln |X|}\f{\pa }{\pa \ln M_j}\]\[\f{\pa \ln M_i}{\pa \ln |X|}\f{\pa }{\pa \ln M_i}\]\ln Z_a(\mu,M_a)~,\nn\\
&=&\[\f{\pa \ln M_j}{\pa \ln |X|}\f{\pa \ln M_i}{\pa \ln |X|}\]\[\f{\pa^2 \ln Z_a(\mu,M_a) }{\pa \ln M_i \pa \ln M_j}\].
\eeqa
\eit

The dependencies of $\ln M_a$ on $\ln |X|$ are in general non-trivial, which depends crucially on the properties of matrices $\la_{ij}$ and $m_{ij}$. We will discuss their expressions in the subsequent sections.

It is well known that in the $d\ra\infty$ limit, the anomaly mediation contributions in the deflect AMSB are sub-leading and the gauge mediation contributions are dominant. So we will derive the  EGM contributions first and return to deflected AMSB cases subsequently.

\section{\label{EGM} Analytical expressions within EGM}
We assume that the mass thresholds of the $N$ messengers can be degenerated and separated into $'p'$ groups as
\beqa
(\underbrace{M_1,\cdots,M_1}_{n_1}, \underbrace{M_2,\cdots,M_2}_{n_2},\cdots,\underbrace{M_p,\cdots,M_p}_{n_p})~.
\eeqa
with $\sum\limits_{i=1}^p n_p=N$.

The gauge coupling at a scale $\mu$ that below all the messenger thresholds are given as
\beqa
\f{1}{g_i^2(\mu, X)}&=&\f{1}{g_i^2(\Lambda)}+\f{b_i^\pr}{8\pi^2}\ln\f{\Lambda}{M_1}
 +\f{b_i^\pr-n_1}{8\pi^2}\ln\f{M_1}{M_2}+\f{b_i^\pr-\sum\limits_{k=1}^2 n_k}{8\pi^2}\ln\f{M_2}{M_3}+\cdots+\f{b_i^\pr-N}{8\pi^2}\ln\f{M_p}{\mu},\nn\\
&=&\f{1}{g_i^2(\Lambda)}+\f{b_i^\pr}{8\pi^2}\ln{\Lambda}-\f{1}{8\pi^2}\ln\det{\cal M}-\f{b_i^\pr-N}{8\pi^2}\ln{\mu}~.
\eeqa
Here we assume that the eigenvalues of the messenger mass matrix are given by $M_1\geq M_2\geq \cdots\geq M_p$.
So we can obtain that gaugino mass
\beqa
M_{i}&=&g_i^2\(\f{F_\phi}{2}\f{\pa}{\pa \ln\mu}-\sum\limits_{a=1}^p\f{d F_\phi}{2}\f{\pa \ln M_a}{\pa \ln |X|}\f{\pa}{\pa \ln M_a}\)\f{1}{g_i^2}(\mu,M_a)~,\nn\\
&=&-F_\phi \f{g_i^2}{16\pi^2}b_i+{d F_\phi}\f{g_i^2}{16\pi^2}\sum\limits_{a=1}^pn_a\f{\pa \ln M_a}{\pa \ln |X|}~,\nn\\
&=&-F_\phi \f{g_i^2}{16\pi^2}b_i+{d F_\phi}\f{g_i^2}{16\pi^2} \f{\pa \ln \det{\cal M}}{\pa \ln |X|}~,\nn\\
&\equiv&-F_\phi\f{g_i^2}{16\pi^2}\[b_i-d~ n_0\].
\eeqa
Here $b_i=b_i^\pr-N$, which are given by
\beqa
(b_1,~b_2,~b_3)=(33/5,~1,~-3)~,
\eeqa
with $N=N_{\bf 5}+3N_{\bf 10}$ and $n_0$ is given in eqn.(\ref{det1}).

For the soft sfermion masses and trilinear couplings, we need the dependence of wavefunction $Z_i$ on the messenger thresholds $M_a(\sqrt{X^\da X})$. The derivative of $\ln Z_i$ with respect to $\ln |X|$ can be obtained via
\beqa
&&\f{\pa \ln Z_l\[\mu,g_i(\mu^\pr), y_i(\mu^\pr), M_a\]}{\pa \ln |X|}\nn\\
 &=&\sum\limits_{M_j}\sum\limits_{g_i(\mu^\pr)}\f{\pa\ln M_j}{\pa \ln |X|}\f{\pa\ln g_i(\mu^\pr)}{\pa \ln M_j} \f{\pa \ln Z_l\[\mu,g_i(\mu^\pr), y_i(\mu^\pr), M_a\]}{\pa\ln g_i(\mu^\pr)}+\[g_i(\mu^\pr)\ra y_a(\mu^\pr)\]\nn\\
 &+&\sum\limits_{M_a}\f{\pa\ln M_a}{\pa \ln |X|}\f{\pa \ln Z_l\[\mu,g_i(\mu^\pr), y_i(\mu^\pr), M_a\]}{\pa\ln M_a}~.
\eeqa
The sum over $g_i(\mu^\pr)$, which depends on the messenger thresholds $M_i$, take the values $g_i(M_1),g_i(M_2),\cdots,g_i(M_p)$.
The second term always vanishes because the anomalous dimension is continuous across the messenger thresholds.

To get the expressions for wavefunction $Z_i$, we need to the following classification
\bit
\item  In the case $b_i^\pr-\sum\limits_{r=1}^k n_r\neq 0$ for all $0\leq k \leq p$, the expression will fall into class A.
\item  In the case $b_i^\pr-\sum\limits_{r=1}^k n_r= 0$ for some $0\leq k \leq p$, the expression will fall into class B.
\eit

\subsection{Class A: Partition without vanishing gauge beta functions }
To obtain $Z_i$, we can construct an invariant by surveying the anomalous dimension of $Z_i$
\beqa
\f{d~\ln Z_i}{dt}=-\f{1}{8\pi^2}\[d_{ij}^k y_{ijk}^2-2C(r)g_i^2\]~,
\eeqa
and solve the differential equation in the basis of $(y_t^2,y_b^2,y_\tau^2,g_3^2,g_2^2,g_1^2)$ with
\beqa
\label{coefficient:wavefunction}
\f{d}{dt} \ln Z_i=\sum\limits_{l=g_3,g_2,g_1} 2\tl{A_l} \f{d \ln g_{l}}{dt}+\sum\limits_{l=y_t,y_b,y_\tau} 2B_l \f{d \ln y_{l}}{dt}~,
\eeqa
at the scale $\Lambda$. The expressions of the wavefunction can be solved \footnote{ For example, see appendix A in Ref.\cite{analytic-mirage} for details.} as
\beqa
Z_l(\mu,M_a)&=&Z_l(\Lambda)\(\f{y_t^2(\mu)}{y_t^2(\Lambda)}\)^{B_t}\(\f{y_b^2(\mu)}{y_b^2(\Lambda)}\)^{B_b}\(\f{y_\tau^2(\mu)}{y_\tau^2(\Lambda)}\)^{B_\tau}\\
&&\prod\limits_{i=1}^3\[\(\f{g_i^2(M_1)}{g_i^2(\Lambda)}\)^{\f{A_i}{b_i^\pr}}\(\f{g_i^2(M_2)}{g_i^2(M_1)}\)^{\f{A_i}{b_i^\pr-n_1}}
\(\f{g_i^2(M_3)}{g_i^2(M_2)}\)^{\f{A_i}{b_i^\pr-\sum\limits_{i=1}^2 b_i}}
\cdots\(\f{g_i^2(\mu)}{g_i^2(M_{p})}\)^{\f{A_i}{b_i^\pr-N}}\].\nn
\eeqa
 with the corresponding coefficients $\tl{A}_i$ in given as
 $\tl{A}_i\equiv A_i/b_i^\pr,A_i/(b_i^\pr-n_1),\cdots$  at the energy interval $[M_i,M_{i+1}]$ ($M_0\equiv \Lambda$), respectively.

  If all the Yukawa terms within the wavefunction $Z_i$ are neglected, the $A_i$ will take the value $4C(r)_i$ with $C(r)_i$ the quadratic Casimir invariant for the superfield $\Phi_i$.

So we have
\small
\beqa
\label{wave}
\ln Z_l(\mu,M_a)&=&\ln Z_l(\Lambda)+\sum\limits_{i}\left\{-\f{A_i}{b_i^\pr}\ln {g_i^2(\Lambda)}+\(\f{A_i}{b_i^\pr}-\f{A_i}{b_i^\pr-n_1}\) \ln {g_i^2(M_1)}
+\(\f{A_i}{b_i^\pr-n_1}-\f{A_i}{b_i^\pr-\sum\limits_{k=1}^2 n_k}\) \ln {g_i^2(M_2)}\right.\nn\\
&+&\left.\(\f{A_i}{b_i^\pr-\sum\limits_{k=1}^2 n_k}-\f{A_i}{b_i^\pr-\sum\limits_{k=1}^3 n_k}\) \ln {g_i^2(M_3)}+\cdots+ \f{A_i}{b_i^\pr-N} \ln {g_i^2(\mu)}\right\}+\cdots~.
\eeqa

From eqn.(\ref{wave}), we get
\small
\beqa
&&\(\f{\pa \ln Z_l(\mu,M_a)}{\pa\ln g_i(\mu^\pr)}\)=2\left( \f{A_i}{b_i^\pr}-\f{A_i}{b_i^\pr-n_1}~,
\f{A_i}{b_i^\pr-n_1}-\f{A_i}{b_i^\pr-\sum\limits_{k=1}^2 n_k}~,
\cdots~,
\f{A_i}{b_i^\pr-\sum\limits_{k=1}^{p-1} n_k}-\f{A_i}{b_i^\pr-N}~,\f{A_i}{b_i^\pr-N}~
\right),\nn\\
\label{dlnzdg}
\eeqa
\normalsize
with the column indices $g_i(M_1),g_i(M_2),\cdots,g_i(M_p),g_i(\mu)$. From the expressions of gauge coupling at scale $\mu$ within each threshold interval, we can obtain an $ (p+1) \tm p$ matrix
\small
\beqa
&&~~~~~~~~~~~~~~~\bea{ccccc}\ln M_1~~~~~~~~& \ln M_2~~~~~~~~~~~~~&\ln M_3~~~~~~~~~~~~~~~&\cdots~~~~~~~&\ln M_p~~\eea\nn\\
\(\f{\pa\ln g_i(\mu^\pr)}{\pa \ln M_j}\)_{\mu^\pr,j}&=&
\f{1}{16\pi^2}\left( \bea{ccccc}b_i^\pr g_i^2(M_1)&0&0&\cdots&0\\
n_1 g_i^2(M_2)&(b_i^\pr-n_1)g_i^2(M_2)&0&\cdots&0\\
n_1 g_i^2(M_3)&n_2 g_i^2(M_3)&(b_i^\pr-\sum\limits_{i=1}^2 n_i)g_i^2(M_3)&\cdots&0\\
\cdots&\cdots&\cdots&\cdots&\cdots\\
n_1 g_i^2(M_p)&n_2g_i^2(M_p)&n_3 g_i^2(M_p)&\cdots&(b_i^\pr-\sum\limits_{i=1}^{p-1} n_i)g_i^2(M_p)\\
n_1 g_i^2(\mu)&n_2 g_i^2(\mu)&n_3 g_i^2(\mu)&\cdots& n_p g_i^2(\mu)\\
\eea\right)\bea{c}g_i(M_1)\\g_i(M_2)\\g_i(M_3)\\\\\cdots\\g_i(M_p)\\g_i(\mu)\eea\nn\\
\label{dgdlnM}
\eeqa
\normalsize
 So we have
\beqa\label{dlnM-nondeg}
&&U_{b}\equiv\left(\f{\pa \ln Z_l(\mu,M_a)}{\pa \ln M_b}\right)^T
\equiv-\sum\limits_{i=1,2,3}\f{A_i}{8\pi^2}U_{b;i}~,\\
 U_{b;i}&=&\(\bea{c}n_1\(P_i[1]+Q_i[1]\)\\n_2\(P_i[2]+Q_i[2]\)\\n_3\(P_i[3]+Q_i[3]\)\\\cdots\\n_p\(P_i[p]+Q_i[p]\)\eea\)
\equiv\(\bea{c}\[\f{n_1 g_i^2(M_1)}{b_i^\pr-n_1}-\f{n_1 g_i^2(\mu)}{b_i^\pr-N}\]
+\sum\limits_{a=2}^p \f{n_1 n_a g_i^2(M_a)}{(b_i^\pr-\sum\limits_{k=1}^{a-1} n_k) (b_i^\pr-\sum\limits_{k=1}^{a} n_k)}\\
\[\f{n_2 g_i^2(M_2)}{b_i^\pr-\sum\limits_{k=1}^{2} n_k}-\f{n_2 g_i^2(\mu)}{b_i^\pr-N}\]
+\sum\limits_{a=3}^p \f{n_2 n_a g_i^2(M_a)}{(b_i^\pr-\sum\limits_{k=1}^{a-1} n_k) (b_i^\pr-\sum\limits_{k=1}^{a} n_k)}\\
\[\f{n_3 g_i^2(M_3)}{b_i^\pr-\sum\limits_{k=1}^{3} n_k}-\f{n_3 g_i^2(\mu)}{b_i^\pr-N}\]
+\sum\limits_{a=4}^p \f{n_3 n_a g_i^2(M_a)}{(b_i^\pr-\sum\limits_{k=1}^{a-1} n_k) (b_i^\pr-\sum\limits_{k=1}^{a} n_k)}\\
\cdots\cdots\\
\[\f{n_p g_i^2(M_p)}{b_i^\pr-N}-\f{n_p g_i^2(\mu)}{b_i^\pr-N}\]
\eea\)\nn
\eeqa
with the column indices corresponding to $M_1,M_2,\cdots,M_p$ etc. Here we rewrite our expressions neatly by define
\beqa
P_i[a]&=&\f{g_i^2(M_a)}{b_i^\pr-\sum\limits_{k=1}^{a} n_k}-\f{g_i^2(\mu)}{b_i^\pr-N}~,\nn\\
Q_i[a]&=&\sum\limits_{c=a+1}^p \f{ n_c g_i^2(M_c)}{(b_i^\pr-\sum\limits_{k=1}^{c-1} n_k) (b_i^\pr-\sum\limits_{k=1}^{c} n_k)}~,
\eeqa
within which $b_i\equiv b_i^\pr-N$ is just the beta function coefficient below all the messenger thresholds. From the previous expressions, we can check that the each row will vanish if we neglect the scale dependence of $g_i^2$. This observation agrees with the ordinary conclusion that the trilinear couplings of GMSB vanish if no Yukawa deflections are present.

From the expressions in eqn.(\ref{dlnM-nondeg}), we can obtain the symmetric matrix (for indices $j$ and $k$)
\beqa
\(\f{\pa^2 \ln Z_a(\mu,M_a) }{\pa \ln M_b \pa \ln M_a}\)^T
&\equiv& -\sum\limits_{i=1,2,3}\f{4 A_i}{(16\pi^2)^2} K_{ab;i}~,
\eeqa
with the contributions $K_{ab;i}$ from each $i=1,2,3$ gauge fields given by
\beqa
\left(
\bea{ccccccc}
~n_1^2 X_i[1],& ~n_1n_2 Y_i[2],&~ n_1 n_3 Y_i[3],&~n_1 n_4 Y_i[4],&~n_1 n_5 Y_i[5],& \cdots, & n_1 n_p Y_i[p]\\
n_1n_2Y_i[2],& ~n_2^2 X_i[2], & ~n_2 n_3 Y_i[3],&~ n_2 n_4 Y_i[4],&~ n_2 n_5 Y_i[5],& \cdots, & n_2n_p Y_i[p] \\
n_1n_3Y_i[3],& ~n_2n_3Y_i[3], &~ n_3^2  X_i[3],&~n_3 n_4 Y_i[4],&~n_3 n_5 Y_i[5],&\cdots&  n_3n_p Y_i[p]\\
n_1n_4Y_i[4],&~ n_2n_4Y_i[4], & ~n_3n_4 Y_i[4],&~~n_4^2 X_i[4],&~~n_4 n_5 Y_i[5],&\cdots&  n_4n_p Y_i[p]\\
n_1n_5Y_i[5],&~ n_2n_5Y_i[5], & ~n_3n_5 Y_i[5],&~~n_4n_5X_i[5],&~~ n^2_5 X_i[5],&\cdots&  n_5n_p Y_i[p]\\
\cdots&\cdots&\cdots&\cdots&&\cdots\\
n_1n_p Y_i[p], &~ n_2n_p Y_i[p],&~n_3 n_p Y_i[p],&~n_4 n_p Y_i[p],&~n_5 n_p Y_i[p],&\cdots,&n_p^2 X_i[p],
\eea\right).
\label{Kmatrix-1}
\eeqa
The functions within $K_{ab;i}$ are defined as
\beqa
X_i[a]&=&G^E_i[a]+K^E_i[a]~,~~~~~Y_i[a]= F_i^E[a]+K_i^E[a]~,
\eeqa
within which
\beqa
F^E_i[a]&=&\f{g_i^4(M_a)}{b_i^\pr-\sum\limits_{i=1}^{a} n_i}-\f{g_i^4(\mu)}{b_i^\pr-N}~,\nn\\
G^E_i[a]&=&\f{( b_i^\pr-\sum\limits_{i=1}^{a-1} n_i) g_i^4(M_a)}{n_a (b_i^\pr-\sum\limits_{i=1}^{a} n_i)}-\f{ g_i^4(\mu)}{b_i^\pr-N}~,\nn\\
K^E_i[a]&=&\sum\limits_{c=a+1}^p \f{ n_c g_i^4(M_c)}{(b_i^\pr-\sum\limits_{i=1}^{c-1} n_i) (b_i^\pr-\sum\limits_{i=1}^{c} n_i)}~,
\eeqa
Here we define the summation to vanish if its index lies out of its definition range. For example, $\sum\limits_{i=1}^0(\cdots)\equiv \sum\limits_{i=p+1}^p(\cdots)=0$.
From the previous expressions, we can check that each non-diagonal element of $K_{ab}$ will vanish if we neglect the scale dependence of $g_i^2$. Only the diagonal elements of $K_{ab}$ can give non-vanishing values of order\beqa\label{approx:AA}  K_{aa}\sim n_a^2 \f{g_i^4}{n_a}=n_a g_i^4. \eeqa

The inclusion of top-Yukawa coupling is straightforward in the analytical expressions. The scale dependence of top-Yukawa in the simplest case, in which only the leading top Yukawa $\al_t\equiv y_t^2/4\pi$ and $\alpha_s\equiv g_3^2/4\pi$ are kept in the anomalous dimension, takes the form
\beqa
\f{d}{dt}\ln \alpha_t&=&\f{1}{\pi}\(3\alpha_t-\f{8}{3}\al_s\)~,~~~~~~~~~~~~~
\f{d}{dt}\ln \al_s=\f{1}{2\pi} b_3\al_s~,
\eeqa
 Define $A=\ln\(\alpha_t \al_s^{\f{16}{3b_3}}\)$, the equation can be written as
\beqa
{d}\[ e^{-A}\]&=&-\f{3}{\pi}\al_s^{-\f{16}{3b_3}} dt=
-\f{6}{b_3}\al_s^{-\f{16}{3b_3}-2}d\al_s~.
\eeqa
So we can exactly solve the differential equation to get
\beqa
\[\f{\alpha_t(\mu)}{\al_t(\Lambda)}\( \f{\al_s(\mu)}{\al_s(\Lambda)}\)^{\f{16}{3b_3}}\]^{-1}
=1-\f{6\al_t(\Lambda)}{\f{16}{3}+b_3}\[\al_s(\Lambda)^{-1}-\(\f{\al_s(\mu)}{\al_s(\Lambda)}\)^{-\f{16}{3b_3}}\al^{-1}_s(\mu)\].\nn
\eeqa

Expanding the expressions and neglect high order terms, we finally have
\beqa
\[\ln {\alpha_t(\mu)}-\ln{\al_t(\Lambda)}\]
&\approx&\[-\f{8}{3\pi}\al_s(\mu)+\f{3}{\pi}\al_t(\mu)\]\ln\(\f{\Lambda}{\mu}\).
\eeqa
 It can be observed that the (leading order approximation) expression within the square bracket is just the beta function of top Yukawa coupling.
As there are no Yukawa deflection contributions related to the introduction of messengers, the Yukawa coupling contributions will not enter the expression within the GMSB part of deflection GM.
\subsection{Class B: Partition with vanishing gauge beta functions }
In previous discussions, apparent poles $A_i/(b_i^\pr-\sum\limits_{a=1}^j n_a)$ may arise in the expressions if the gauge beta function coefficient $b_i^\pr-\sum\limits_{a=1}^j n_a=0$  between certain messenger scales.  For example, with $N> 3$ non-degenerate messengers in ${\bf 5}\oplus \bar{\bf 5}$ representation, the beta function for $\alpha_3$, which is given by $b_3^\pr= -3+N$, will vanish after decoupling $N-3$ family of vector-like messengers at one-loop level. (The beta function for $i=1,2$ gauge fields will not encounter this possibility.) Such an artificial pole can be resolved by revisiting the deduction procedure of wavefunctions.

Assume that all the $'i'$-th gauge coupling beta function coefficients $b_i^\pr-\sum\limits_{a=1}^{k-1} n_a$ are non-vanishing for $k < j $. After integrating out the $n_j$ family of vector-like messengers at $M_j$ scale, the beta function coefficient is assumed to vanish (so that $b_i^\pr-\sum\limits_{a=1}^{j} n_a=0$).

The wavefunction at the $M_j$ scale takes value
\beqa
Z_l(M_j,M_a)&=&Z_l(\Lambda)\prod\limits_{i=1}^3\[\(\f{g_i^2(M_1)}{g_i^2(\Lambda)}\)^{A_i/b_i^\pr}\(\f{g_i^2(M_2)}{g_i^2(M_1)}\)^{A_i/(b_i^\pr-n_1)}
\cdots\(\f{g_i^2(M_j)}{g_i^2(M_{j-1})}\)^{A_i/(b_i^\pr-\sum\limits_{a=1}^{j-1} n_a)}\]~,\nn\\
&&\(\f{y_t^2(M_j)}{y_t^2(\Lambda)}\)^{B_t} \(\f{y_b^2(M_j)}{y_b^2(\Lambda)}\)^{B_b}\(\f{y_{\tau}^2(M_j)}{y_{\tau}^2(\Lambda)}\)^{B_\tau}~.
\eeqa
As the $'i'$-th beta function vanishes at one-loop level
\beqa
\f{d g_i}{dt}=0~,~~~~~~~ M_{j+1}<\mu<M_{j}~
\eeqa
it can be seen as a constant between $M_{j+1}<\mu<M_{j}$. Within this range, the RGE invariant became
\beqa
\f{d}{dt}\ln \[ Z(\mu)\prod\limits_{l=y_t,y_b,y_\tau}\[y_l(\mu)\]^{-2B_l}\prod\limits_{k\neq i}\[g_k(\mu)\]^{-2\tl{A}_k}\]=D_i g_i^2(\mu)=D_i g_i^2(M_{j})~,
\eeqa
and we can deduce that
 \beqa
&& Z(\mu)\prod\limits_{l=y_t,y_b,y_\tau}\[y_l(\mu)\]^{-2B_l}\prod\limits_{k\neq i}\[g_k(\mu)\]^{-2\tl{A}_k}\nn\\
 &=&\(\f{\mu}{M_{j}}\)^{D_i g_i^2(M_j)}\[Z(M_{j})\prod\limits_{l=y_t,y_b,y_\tau}\[y_l(M_{j})\]^{-2B_l}\prod\limits_{k\neq i}\[g_k(M_{j})\]^{-2\tl{A}_k}\].
\eeqa
The value $D_i\equiv A_i/8\pi^2$ with the value $A_i$ given in the appendix A. We keep to use $'D_i'$ in this paper to indicate clearly the consequence of vanishing one-loop beta functions.

So, for $M_{j+1}<\mu<M_{j}$
\small
\beqa
Z(\mu)&=& Z(M_{j})\(\f{\mu}{M_{j}}\)^{D_i g_i^2(M_j)}\prod \limits_{l=y_t,y_b,y_\tau}\[\f{y_l(M_{j})}{y_l(\mu)}\]^{-2B_l}\prod\limits_{k\neq i}\[\f{g_k(M_{j})}{g_k(\mu)}\]^{-2\tl{A}_k}\nn\\
&=& Z(\Lambda)\prod \limits_{l=y_t,y_b,y_\tau}\[\f{y_l(\Lambda)}{y_l(\mu)}\]^{-2B_l} \prod\limits_{k\neq i}\[\(\f{g_k^2(M_1)}{g_k^2(\Lambda)}\)^{\f{A_k}{b_k^\pr}}\(\f{g_k^2(M_2)}{g_k^2(M_1)}\)^{\f{A_k}{b_k^\pr-n_1}}
\cdots\(\f{g_k^2(\mu)}{g_k^2(M_{j})}\)^{\f{A_k}{b_k^\pr-\sum\limits_{a=1}^{j}{n_a}}}\],\nn\\
&&\[\(\f{g_i^2(M_1)}{g_i^2(\Lambda)}\)^{\f{A_i}{b_i^\pr}}
\(\f{g_i^2(M_2)}{g_i^2(M_1)}\)^{\f{A_i}{b_i^\pr-n_1}}
\cdots\(\f{g_i^2(M_{j})}{g_i^2(M_{j-1})}\)^{\f{A_i}{b_i^\pr-\sum\limits_{a=1}^{j-1}{n_a}}}\]\(\f{\mu}{M_{j}}\)^{D_i g_i^2(M_{j})}.\nn
\eeqa
\normalsize
and for $ \mu<M_{p}$
\small
\beqa
Z(\mu)&=& Z(\Lambda) \prod \limits_{l=y_t,y_b,y_\tau}\[\f{y_l(\Lambda)}{y_l(\mu)}\]^{-2B_l}\nn\\
&&\prod\limits_{k\neq i}\[\(\f{g_k^2(M_1)}{g_k^2(\Lambda)}\)^{\f{A_k}{b_k^\pr}}\(\f{g_k^2(M_2)}{g_k^2(M_1)}\)^{\f{A_k}{b_k^\pr-n_1}}
\cdots\(\f{g_k^2(M_j)}{g_k^2(M_{j-1})}\)^{\f{A_k}{b_k^\pr-\sum\limits_{a=1}^{j}{n_a}}}\cdots
\(\f{g_k^2(\mu)}{g_k^2(M_{p})}\)^{\f{A_k}{b_k}}\],\nn\\
&&\[\(\f{g_i^2(M_1)}{g_i^2(\Lambda)}\)^{\f{A_i}{b_i^\pr}}\(\f{g_i^2(M_2)}{g_i^2(M_1)}\)^{\f{A_i}{b_i^\pr-n_1}}
\cdots\(\f{g_i^2(M_{j})}{g_i^2(M_{j-1})}\)^{\f{A_i}{b_i^\pr-\sum\limits_{a=1}^{j-1}{n_a}}}\]\(\f{M_{j+1}}{M_{j}}\)^{D_i g_i^2(M_{j})}\nn\\
&&\(\f{g_i^2(M_{j+2})}{g_i^2(M_{j+1})}\)^{\f{A_i}{b_i^\pr-\sum\limits_{a=1}^{j+1}{n_a}}}\cdots \(\f{g_i^2(\mu)}{g_i^2(M_p)}\)^{\f{A_i}{b_i}}.
\eeqa
\normalsize
For $ \mu<M_{p}$,
\beqa
\f{d \ln Z\[\mu,g_i(\mu^\pr),M_j,M_{j+1}\]}{d \ln M_a}=\[\sum\limits_{g_i(\mu^\pr)}\f{\pa g_i(\mu^\pr)}{\pa \ln M_a}\f{\pa }{\pa g_i(\mu^\pr)}+\f{\pa }{\pa \ln M_a}\]\ln Z\[\mu,g_i(\mu^\pr),M_j,M_{j+1}\]~.\nn
\eeqa
with the last term gives non-vanishing contributions only for $a=j,j+1$
\beqa
\f{\pa }{\pa \ln M_a} \ln Z\[\mu,g_i(\mu^\pr),M_j,M_{j+1}\]=\[\delta_{a,j+1}-\delta_{a;j}\] D_i g_i^2(M_{j})~,
\eeqa
 From the general expressions, we can see that the $'j'$-th and $'(j+1)'$-th components will change into
\beqa
&&\(\f{\pa \ln Z[\mu; g_i(\mu^\pr), M_a]}{\pa \ln g_i(\mu^\pr)}\)_{j,j+1}=2\left( \f{A_i}{b_i^\pr-\sum\limits_{a=1}^{j-1} n_a}+D_i g_i^2(M_j)\ln\f{M_{j+1}}{M_{j}}~,
-\f{A_i}{b_i^\pr-\sum\limits_{a=1}^{j+1} n_a}~\right),\nn
\eeqa
while other columns are unchanged as eqn.(\ref{dlnzdg}) if $b_i(\mu)=0$ for $M_{j+1}<\mu<M_{j}$. The matrix $\pa \ln g_i(\mu^\pr)/\pa M_j$, which is given by eqn.(\ref{dgdlnM}), is unchanged. Then the contributions from the $'i'$-th gauge field, which has vanishing beta functions between $[M_{j+1},M_j]$, are given as
\beqa
U_{b;i}&\equiv&\left.\f{d \ln Z[g_i(\mu^\pr),M_n]}{d \ln M_b}\right|_{i}\\
&=&\[\f{\pa g_i(\mu^\pr)}{\pa \ln M_b}\f{\pa }{\pa g_i(\mu^\pr)}+\f{\pa }{\pa \ln M_b}\]\ln Z\[g_i(\mu^\pr),M_j,M_{j+1}\]\nn\\
&=&-\f{A_i}{8\pi^2}\(\bea{c} n_1 \(P_i^S[1]+Q_i^S[1] -\f{D_i}{A_i}  g_i^4(M_j)\ln\f{M_{j+1}}{M_{j}}\)~,\\
 n_2 \(P_i^S[2]+Q_i^S[2] -\f{D_i}{A_i}  g_i^4(M_j)\ln\f{M_{j+1}}{M_{j}}\)~,\\
 \cdots\\
 n_j\(P_i^S[j]+Q_i^S[j]-\f{D_i}{A_i} g_i^4(M_j)\ln\f{M_{j+1}}{M_{j}}\)+\f{8\pi^2}{A_i}D_ig_i^2(M_j)~,\\
 n_{j+1}\( P_i^S[j+1]+Q_i^S[j+1]\)-\f{8\pi^2}{A_i}D_ig_i^2(M_j)~,\\
 n_{j+2}\( P_i^S[j+2]+Q_i^S[j+2]\)~,\\
  \cdots\\
 n_p\( P_i^S[p]+Q_i^S[p]\])~.
\eea\).\nn
\label{trilinear}
\eeqa

Within the deduction, we use the fact that $g_i(M_j)=g_i(M_{j+1})$ and $b_i^\pr-\sum\limits_{k=1}^j n_k=0$. Besides, we define within the expression
\beqa
Q_i^S[c]&=&\left\{\bea{c}\sum\limits_{a=c+1;a\neq j,j+1}^p \f{ n_a g_i^2(M_a)}{(b_i^\pr-\sum\limits_{k=1}^{a-1} n_k) (b_i^\pr-\sum\limits_{k=1}^{a} n_k)}
+\f{(n_j+n_{j+1}) g_i^2(M_j)}{(b_i^\pr-\sum\limits_{k=1}^{j-1} n_k)(b_i^\pr-\sum\limits_{k=1}^{j+1} n_k)}~,~~~~~~1\leq c \leq j\nn\\
  \nn\\
\sum\limits_{a=c+1}^p \f{ n_a g_i^2(M_a)}{(b_i^\pr-\sum\limits_{k=1}^{a-1} n_k) (b_i^\pr-\sum\limits_{k=1}^{a} n_k)}~,~~~~~~~~~~~~~~~~~~~~~j+1\leq c\leq p
\eea\right.\\
P_i^S[c]&=&\left\{\bea{c}
\f{g_i^2(M_{c})}{b_i^\pr-\sum\limits_{k=1}^{c} n_k}-\f{ g_i^2(\mu)}{b_i^\pr-N}
~,~~~~~~~~~~c\neq j,j+1\\
-\f{g_i^2(\mu)}{b_i^\pr-N} ~,~~~~~~~c=j,j+1\\
\eea\right.
\eeqa
Within the expression
\beqa
\f{(n_j+n_{j+1}) g_i^2(M_j)}{(b_i^\pr-\sum\limits_{k=1}^{j-1} n_k)(b_i^\pr-\sum\limits_{k=1}^{j+1} n_k)}
=-\f{n_j+n_{j+1}}{n_j n_{j+1}}g_i^2(M_j)~.
\eeqa
We note that when $c$ takes value $j-1$ or $j$ in the summation of $Q_i^S[c]$, the sum skip $j,j+1$ and begins at $a=j+2$.

 From the previous expressions, we can see that the each row will vanish if we neglect the scale dependence of $g_i^2$ and higher order $g_i^6$ terms. In fact, with such an approximation, the $j$-th and $j+1$-th row is given by \beqa U_{b;j}\sim -\f{g^2}{n_j} n_j+g^2\sim 0~,\nn\\ U_{b;j+1}\sim \f{g^2}{n_{j+1}} n_{j+1}-g^2\sim 0~. \eeqa
 

In the summation
 \beqa
  \f{\pa \ln Z[\mu,g_i(\mu^\pr),M_n]}{\pa \ln M_b}=
  \sum\limits_{i=1}^3 \left.\f{\pa \ln Z[\mu,g_i(\mu^\pr),M_n]}{\pa \ln M_b}\right|_i~,
 \eeqa
 the expressions for $i=1,2$ gauge fields (which have no vanishing beta functions) are still given by
 \beqa
\left.\left(\f{\pa \ln Z[\mu,g_i(\mu^\pr)]}{\pa \ln M_b}\right)\right|_{i={1,2}}
\equiv-\f{A_i}{8\pi^2}\(\bea{c}n_1\(P_i[1]+Q_i[1]\)\\n_2\(P_i[2]+Q_i[2]\)\\n_3\(P_i[3]+Q_i[3]\)\\\cdots\\n_p\(P_i[p]+Q_i[p]\)\eea\)~,
\eeqa
 from eqn.(\ref{dlnM-nondeg}).

With previous results, we can derive the expression of $\f{\pa^2}{\pa \ln M_a\ln M_b} Z[\mu; g_i(\mu^\pr), M_n]$ from $'i'$-th ( here $i=3$) gauge fields
\beqa
&&\f{\pa}{\pa \ln M_a}\(\left.\f{\pa \ln Z[g_i(\mu^\pr),M_n]}{\pa \ln M_b}\right|_i\)\equiv -\f{A_i}{8\pi^2}\[\f{\pa \ln g_j(\mu^\pr)}{\pa \ln M_a}\f{\pa}{\pa \ln g_j(\mu^\pr)}+\delta_{a;j,j+1}\f{\pa}{\pa M_{j,j+1}}\] V_{b;i}\nn\\
&=&-\f{4 A_i}{(16\pi^2)^2}K_{ab;i}
\eeqa
with $K_{ab;i}$ a symmetric matrix given as
\scriptsize
\beqa\left(
\bea{ccccccccc}
n_1^2 J[1]&n_1 n_2 H[2]&n_1n_3H[3]&\cdots& n_1 n_j H[j]&n_1 n_{j+1}H[j+1]&n_1 n_{j+2}H[j+2]&\cdots&n_1n_p H[p]  \\
n_1 n_2 H[2]&n_2^2 J[2]&n_2n_3H[3]&\cdots& n_2 n_j H[j]&n_2n_{j+1}H[j+1]&n_2n_{j+2}H[j+2]&\cdots&n_2n_p H[p]\\
n_1 n_3 H[3]&n_3 n_2 H[3]&n_3^2 J[3]&\cdots& n_3 n_j H[j]&n_3n_{j+1}H[j+1]&n_3n_{j+2}H[j+2]&\cdots &n_3n_p H[p]\\
\cdots&\cdots&\cdots&\cdots&\cdots&\cdots&\cdots&\cdots&\cdots\\
n_1 n_j H[j]&n_2 n_j H[j]&n_3 n_j H[j]&\cdots&n_j^2J[j]&n_j n_{j+1}H[j+1]&n_j n_{j+2}H[j+2]&\cdots&n_j n_p H[p]\\
n_1 n_{j+1}H[j+1] &n_2n_{j+1}H[j+1]&n_3n_{j+1}H[j+1]&\cdots&n_j n_{j+1}H[j+1]&n_{j+1}^2J[j+1]&n_{j+1} n_{j+2}H[j+2]&\cdots&n_{j+1}n_p H[p]\\
n_1 n_{j+2}H[j+2]&n_2n_{j+2}H[j+2]&n_3n_{j+2}H[j+2]&\cdots&n_j n_{j+2}H[j+2]&n_{j+1} n_{j+2}H[j+2]&n_{j+2}^2 J[j+2]&\cdots&n_{j+2}n_p H[p]\\
\cdots&\cdots&\cdots&\cdots&\cdots&\cdots&\cdots&\cdots&\cdots\\
n_1 n_p H[p]&n_2n_p H[p]&n_3n_p H[p]&\cdots&n_j n_p H[p]&n_{j+1} n_p H[p]&n_{j+2} n_p H[p]&\cdots& n_p^2 J[p]\\
\eea\right)\nn~,\\
\label{scalarmass}
\eeqa
\normalsize
The functions within $K_{ab;i}$ are defined as
\beqa
J[m]&=&\left\{\bea{c} G_i^S[m]+K_i^S[m]-2\f{D_i}{A_i} g_i^6 (M_j)\ln\f{M_{j+1}}{M_{j}}~,~~~~~~~~~~~~~~~~~~~~~~~~~~~~1 \leq m\leq j-1\\
G_i^S[j]+K_i^S[j]+16\pi^2\f{D_i}{n_j A_i}  g_i^4(M_j)-2\f{D_i}{A_i}  g_i^6(M_j)\ln\f{M_{j+1}}{M_{j}}~,~~~m=j~\nn\\
G_i^S[m]+K_i^S[m]~,~~~~~~~~~~~~~~~~~~~~~~~~~~~~~~~~~~~~~~~~~~~~~~ j+1 \leq m \leq p \eea\right.\\
H[m]&=&\left\{\bea{c} F_i^S[m]+K_i^S[m]-2\f{D_i}{A_i} g_i^6 (M_j)\ln\f{M_{j+1}}{M_{j}}~,~~~~~~~~~~~~~~~~~~~~~~~~~~~1 \leq m\leq j-1~\nn\\
F_i^S[j]+K_i^S[j+1]+8\pi^2\f{ D_i}{n_j A_i}  g_i^4(M_j)-2\f{D_i}{A_i}  g_i^6(M_j)\ln\f{M_{j+1}}{M_{j}}~,~~~m=j~\nn \\
F_i^S[j+1]+K_i^S[j+1]-8\pi^2\f{D_i}{n_{j+1}A_i}  g_i^4(M_j)~,~~~~~~~~~~~~~~~~~~~~~~~~~~~m=j+1~\nn\\
F_i^S[m]+K_i^S[m]~,~~~~~~~~~~~~~~~~~~~~~~~~~~~~~~~~~~~~~~~~~~~~~~~ j+2 \leq m \leq p
\eea\right.
\eeqa
with
\beqa
K_i^S[c]&=&\left\{\bea{c}\sum\limits_{a=c+1;a\neq j,j+1}^p \f{ n_a g_i^4(M_a)}{(b_i^\pr-\sum\limits_{k=1}^{a-1} n_k) (b_i^\pr-\sum\limits_{k=1}^{a} n_k)}
+\f{(n_j+n_{j+1}) g_i^4(M_j)}{(b_i^\pr-\sum\limits_{k=1}^{j-1} n_k)(b_i^\pr-\sum\limits_{k=1}^{j+1} n_k)}~,~~~~1\leq c \leq j \nn\\
\sum\limits_{a=c+1}^p \f{ n_a g_i^4(M_a)}{(b_i^\pr-\sum\limits_{k=1}^{a-1} n_k) (b_i^\pr-\sum\limits_{k=1}^{a} n_k)}~,~~~~~~~~~~~~~~~~~~~~~~~~~~~~~~~~~~~~~j+1\leq c\leq p
\eea\right.\\
F_i^S[c]&=&\left\{\bea{c}\f{g_i^4(M_{c})}{b_i^\pr-\sum\limits_{k=1}^{c} n_k}-\f{ g_i^4(\mu)}{b_i^\pr-N}
~,~~~~~~c\neq j,j+1\\
\f{n_j+n_{j+1}}{b_i^\pr-\sum\limits_{k=1}^{j-1} n_k}\f{g_i^4(M_{c})}{b_i^\pr-\sum\limits_{k=1}^{j+1} n_k}-\f{ g_i^4(\mu)}{b_i^\pr-N}~,~~~~~~c=j\\
-\f{g_i^4(\mu)}{b_i^\pr-N} ~,~~~~~~~~~~~~c=j+1 \\
\eea\right.\\
G_i^S[c]&=&\left\{
\bea{c}
\f{(b_i^\pr-\sum\limits_{k=1}^{c-1} n_k) g_i^4(M_a)}{n_c (b_i^\pr-\sum\limits_{k=1}^{c} n_k)}-\f{g_i^4(\mu)}{b_i^\pr-N}~,~~c\neq j,j+1\\
-\f{g_i^4(\mu)}{b_i^\pr-N} ~,~~~~~~~~~~~~c=j,j+1
\eea\right.
\eeqa

 From the previous expressions, we can check that each non-diagonal element of $K_{ab}(a,b\neq j,j+1)$ will vanish if we neglect the scale dependence of $g_i^2$ and higher order $g_i^6$ terms. The diagonal elements of $K_{ab}(a,b\neq j,j+1)$ can give non-vanishing values of order \beqa  K_{aa}\equiv n_a^2 J[a]\sim n_a^2\f{g_i^4}{n_a}=n_a g_i^4~. \eeqa  The $J[j]$ and $H[j]$ term will be given as \beqa J[j]&\sim &  -\f{g_i^4}{n_{j}}+ \f{2}{n_j}g_i^4~,~~~~ J[j+1] \sim \f{g_i^4}{n_{j+1}}~,\nn\\ H[j]&\sim &  -\f{g_i^4}{n_{j}}+ \f{g_i^4}{n_j}\sim 0~,~~~~H[j+1] \sim   \f{g_i^4}{n_{j+1}}-\f{g_i^4}{n_{j+1}}\sim 0~, \eeqa if we neglect the scale dependence of $g_i^2$ and higher order $g_i^6$ terms. 

The contributions from $i=1,2$ gauge fields are still given by eqn.(\ref{Kmatrix-1}). The total contributions are given by the sum of $i=1,2,3$ gauge fields.
\subsection{Dependence of $\ln M_a$ on $\ln X$}

 As noted before, with non-trivial $U(1)_R$ symmetry, the messenger determinant is proven by \cite{EGM} to be a monomial in $X$
\beqa
\label{det}
\det \(\la_{ij}X+\ka_{ij}T\)=X^{n_0} G(\la, \ka T)~,
\eeqa
with $\ka_{ij} \langle T\rangle=m_{ij}$. We need to replace the lowest component VEV $\langle X\rangle\equiv M$ of $X$, which should be substituted in ${\cal M}$ to get all the eigenvalues, by its superspace extension $\sqrt{X^\da X}$ into ${\cal M}$ to incorporate the SUSY breaking effects.

Knowing the value of the determinant, it is still nontrivial to express the eigenvalues of ${\cal M}$ in terms of $\langle X\rangle$.
 Fortunately, the asymptotic behavior will display a simple form.
In large $\langle X\rangle$ region, $r_\la\equiv rank(\la_{ij})$ messengers acquire masses  ${\cal O}(\langle X\rangle)$ while the remaining  $N-r_\la$ messengers acquire masses of order
\beqa
M_i\sim \f{m^{n_i+1}}{\langle X\rangle^{n_i}}~,~~~~~~~ \sum\limits_{i=1}^{N-r_\la} n_i=r_\la-n_0~,
\eeqa
with $n_i\geq 0$. At small $\langle X\rangle$ region, $r_m\equiv rank(m_{ij})$ messengers acquire masses ${\cal O}(m)$ while the remaining $N-r_m$ messengers acquire masses of order
\beqa
M_i\sim  \f{\langle X\rangle^{\tl{n}_i+1}}{m^{\tl{n}_i}}~,~~~~~~~ \sum\limits_{i=1}^{N-r_m} (\tl{n}_i+1)=n_0~.
\eeqa
with $\tl{n}_i\geq 0$.

Depending on the singularity properties of the messenger mass matrix, we have the following the discussions

\bit
\item  Type I: $\det m\neq 0$.

In the basis in which $m$ is diagonal, it can easily obtain \cite{EGM} that the eqn.(\ref{det}) takes the form
\beqa
n=0~,~~~~~~\det(\la \langle X\rangle+m )=\det m~,
\eeqa
which necessarily imply $\det \lambda=0$. As the matrix is upper triangular, the eigenvalues are  $m_{ii}$ that do not depend on $\langle X\rangle$.

So in this case, we will have
\beqa
\f{\pa \ln M_i}{\pa \ln |X|}\equiv 0.
\eeqa
So we can see that the gauginos, the trilinear couplings as well as the sfermions will not receive any gauge mediation contributions.

\item Type II: $\det \la\neq 0$.

Similarly, we can obtain an upper triangular matrix with eigenvalues the diagonal elements of diagonalized matrix $\la^\pr_{ii}$. The determinant is
\beqa
n=N~,~~~~~~ \det(\la \langle X\rangle+m )=\langle X\rangle^N \det\la~.
\eeqa
So the eigenvalues will be
$\la^\pr_{ii} \langle X\rangle$ and depends linearly on $\langle X\rangle$. We will arrange $\la^\pr_{ii}$ to obtain the eigenvalues $\tl{M}_1(\langle X\rangle),\tl{M}_2(\langle X\rangle),\cdots,\tl{M}_N(\langle X\rangle)$.

Suppose the $T_i\equiv \la^\pr_{ii}$ are ordered so as that $T_1\geq T_2\geq T_3\cdots \geq T_N$, we define
\beqa
V_i\equiv\f{d \ln M_i}{d\ln |X|}\equiv (1~,1~,1~,\cdots,~1).
\eeqa

For degenerate eigenvalues
\beqa
&&\tl{M}_1=\tl{M}_2=\cdots=\tl{M}_{n_1}\equiv M_1={\lambda}^\pr_{n_1n_1} \langle X\rangle,~~\tl{M}_{n_1+1}=\tl{M}_{n_1+2}=\cdots=\tl{M}_{n_2}\equiv M_2={\lambda}^\pr_{n_2n_2} \langle X\rangle,\nn\\
&&\tl{M}_{n_2+1}=\tl{M}_{n_2+2}=\cdots=\tl{M}_{n_3}\equiv M_3={\lambda}^\pr_{n_3n_3} \langle X\rangle,~\cdots~\nn\\ &&\tl{M}_{n_{p-1}+1}=\tl{M}_{n_{p-1}+2}=\cdots=\tl{M}_{n_p}\equiv M_p={\lambda}^\pr_{n_pn_p} \langle X\rangle,
\eeqa
with $M_1\geq M_2\geq\cdots\geq M_p$, the matrix $V_i$ reduces to a $1\times p$ matrix
\beqa
V_i\equiv\f{d \ln M_i}{d\ln |X|}\equiv (1~,1~,\cdots,~1).
\eeqa
So the soft SUSY breaking parameters from GMSB
\bit
\item The gaugino mass:
 \beqa
 M_i=\f{F_X}{M}\f{g_i^2}{16\pi^2}n_0~.
 \eeqa

\item GMSB contributions to trilinear terms:
\beqa
A_i=\f{A_{ijk}}{y_{ijk}}=\sum\limits_{i,j,k}\f{F_X}{2M}\f{\pa \ln Z_l(\mu,M_a)}{\pa \ln |X|}&=&\f{F_X}{2M}\sum\limits_{i,j,k}\sum\limits_{v=1}^N U_{b} V_b.~
\eeqa

\item Pure GMSB contributions to soft sfermion masses:
\beqa
&&m_i^2=-\f{F_X^2}{4M^2}\f{\pa^2}{\pa \ln |X|^2}\ln Z_i(\mu,M_a)\nn\\
&=&
\f{F_X^2}{4M^2}\sum\limits_{i=1,2,3}\f{4 A_i}{(16\pi^2)^2}\sum\limits_{a,b} V_a K_{ab;i}V_b~.
\eeqa
\eit

 We should note that if hierarchical structure appears within the diagonal elements of the diagonalized  matrix $\la^\pr_{ii}(i=1,\cdots,N)$, the splitting of the messenger scales will lead to large GMSB contributions.

\item Type III: $\det m=\det\la=0$.


As the matrix $\la X+ m$ is non-singular, its eigenvalues can be written as $x_1,\cdots,x_n$ which
should satisfy
\beqa
\prod\limits_{i}x_i=\det(\la X+ m)=X^{n_0} G(\la,m)~,
\eeqa
and
\beqa
\label{trace}
\sum\limits_{i}x_i=-Tr (\la X+ m)=c X+d~.
\eeqa
In the large $X$ region in which $m_{ij}$ can be neglected, we can use linear transformation to put $\la_{ij}$ into
\beqa
\(\bea{cccccc}a_1&&&&&\\&a_2&&&&\\&&\cdots&&&\\&&&a_{r_\la}&\\&&&&0& \\&&&&& \ddots\eea\)
\label{typeIII:X}
\eeqa
There are $r_\la$ messengers with mass of order $X$. As the trace depends linearly on $X$, such $r_\la$ messengers had to be linearly depends on $X$. The remaining messengers can only proportional to an inverse power of $X$ or be a constant.
From the trace, which contains only the constant and the linear $X$ term, the term with negative power of $X$ should appear in pairs or vanish. As the eigenvalues, which contain non-vanishing negative $n_i$ powers, are suppressed by an additional $(m/\langle X\rangle)^{n_i}$ factor, they need to be the lighter eigenvalues.

As $r_\la$ messengers depends linearly on $X$, we can approximately use
\beqa
\f{\pa \ln M_a}{\pa \ln |X|}\equiv  V_a\approx (\underbrace{1,1,\cdots,1}_{r_\la}, \underbrace{0,0,\cdots,-n_{i_1},-n_{i_1},\cdots,-n_{i_k},-n_{i_k}}_{N-r_\la})
\eeqa
with
\beqa
\sum\limits_{k} 2n_{i_k}=r_\la-n_0~.
\eeqa

For degenerate eigenvalues
\beqa
&&\tl{M}_1=\tl{M}_2=\cdots=\tl{M}_{n_1}\equiv M_1=a_{n_1} \langle X\rangle~,~~~\cdots~\nn\\
&&\tl{M}_{n_{k-1}+1}=\tl{M}_{n_{k-1}+2}=\cdots=\tl{M}_{n_{k}}\equiv M_k=a_{n_k} \langle X\rangle,~\nn\\
&&~\tl{M}_{n_k+1}=\tl{M}_{n_k+2}=\cdots=\tl{M}_{n_{k+1}}\equiv M_{k+1}=c_k,~~\cdots~\nn\\
&& \tl{M}_{n_x+1}=\tl{M}_{n_x+2}=\cdots=\tl{M}_{n_{x+1}}\equiv M_{x+1}=c_{x},\nn\\
&& \tl{M}_{n_{x+1}+1}=\tl{M}_{n_{x+1}+2}=\cdots=\tl{M}_{n_{x+2}}\equiv M_{x+2}=b_{n_{x+2}} \langle X\rangle^{-\la_{x+2}},~~~\cdots\nn\\
 &&\tl{M}_{n_{p-1}+1}=\tl{M}_{n_{p-1}+2}=\cdots=\tl{M}_{n_p}\equiv M_p=b_{n_p} \langle X\rangle^{-\la_p},
\eeqa
with $c_k,\cdots,c_x$ some constants eigenvalues of $\la \langle X\rangle+ m$ independent of $X$.
Assume $M_1\geq M_2\geq\cdots\geq M_p$, the matrix $V_i$ reduces to a $1\times p$ matrix
\beqa\label{vv}
\f{\pa \ln M_a}{\pa \ln |X|}\equiv  V_a\approx ( \underbrace{1,\cdots,1}_{k}, \underbrace{0,\cdots,0}_{x-k+1},\underbrace{-\la_{x+2},\cdots,-\la_p}),
\eeqa
with
\beqa
\sum\limits_{i=1}^k n_{i}=r_\la~,~~\sum\limits_{k=2}^{p-x}\(n_{x+k}-n_{x+k-1}\) \la_{x+k}=r_\la-n~.
\eeqa
So we can obtain the GMSB contributions
\beqa
A_i&=&\f{A_{ijk}}{y_{ijk}}=\sum\limits_{i,j,k}\f{F_X}{2M}\f{\pa \ln Z_l(\mu,M_a)}{\pa \ln |X|}=\f{F_X}{2M}\sum\limits_{i,j,k}\sum\limits_{v=1}^p U_{b} V_b~,\nn\\
m_i^2&=&-\f{F_X^2}{4M^2}\f{\pa^2}{\pa \ln |X|^2}\ln Z_i(\mu,M_a)\nn\\
&=&-\f{F_X^2}{4M^2}\sum\limits_{i=1,2,3}\f{4 A_i}{(16\pi^2)^2}\sum\limits_{a,b} V_a K_{ab;i}V_b~.
\eeqa
with the $V_a$ takes the value in eqn.(\ref{vv}).
The partition of $N$ can be obtained numerically by diagonalizing $\la \langle X\rangle+m$ to obtain its eigenvalues as functions of $\langle X \rangle$.


\eit

 The inclusion of EGM in deflected AMSB is straightforward. The AMSB type contributions can be given as
 \beqa
 \f{\pa}{\pa \ln\mu}\ln \[Z_i(\mu,M_a)\]&=&-\f{1}{8\pi^2} G_i^-[g_l(\mu),y_l(\mu)],\nn\\
 \f{\pa^2}{\pa (\ln\mu)^2}\ln \[Z_i(\mu,M_a)\]&=&-\f{1}{8\pi^2}
 \[\f{\pa g_l(\mu)}{\pa \ln \mu}\f{\pa}{\pa g_l(\mu)}+\f{\pa y_l(\mu)}{\pa \ln \mu}\f{\pa}{\pa y_l(\mu)}\]G_i^-[g_l(\mu),y_l(\mu)]\nn\\
 &=&-\f{2}{(16\pi^2)^2}\[\beta_{g_l}\f{\pa}{\pa g_l(\mu)}+\beta_{y_l}\f{\pa}{\pa y_l(\mu)}\]G_i^-[g_l(\mu),y_l(\mu)]~.
 \eeqa
 The interference terms between AMSB and GMSB can also easily obtained with eqn.(\ref{dlnM-nondeg}) and eqn.(\ref{trilinear}).



In ref.(\ref{EGM}), the $'\textit{effective messenger number}'$ is  defined as
\beqa
N_{eff}\equiv \f{\Lambda_G^2}{\Lambda_S^2}~,
\eeqa
with
\beqa
M_i=\f{g_i^2}{16\pi^2}\Lambda_G~,~~~
~m_{\tl{f}}^2=2\f{g_i^4}{(16\pi^2)^2}\sum\limits_{i}C_{\tl{f}}(r)\Lambda_S^2~.
\eeqa
So the approximate value of $N_{eff}$ can be given as
\beqa
N_{eff}=\f{n_0^2 g_i^4}{\sum\limits_{a,b} V_a K_{ab;i}V_b},
\eeqa
by neglecting the scale dependence of $g_i$ and higher order terms in the expressions of soft SUSY parameters. With previous approximation, the value of $N_{eff}$ in Type II can be calculated to be $N_{eff}=N$ after simplifications.  While in Type III EGM, it can be calculated to be $N_{eff}=n_0$. Such a result holds for both Class A and Class B. Taking into account the scale dependence of $g_i$, $N_{eff}$ can be different to $n_0$ and $N$.

\section{\label{Landau}Messengers on GUT and Landau Pole}
We must ensure that no Landau pole will be reached below the GUT scale. It is obvious that the gauge coupling unification will be preserved because the messengers are fitted into complete SU(5) representations.
The presence of (complete GUT representation) messenger fields at an intermediate scale does not modify the value of $M_{GUT}$.
However, proton decay could possibly set constraints on the gauge couplings at the GUT scale.

We can define the quantity
\beqa
\delta=-\sum\limits_{r=1}^p \f{n_r}{2\pi}\ln \f{M_{GUT}}{M_{r}}~,
\eeqa
which contributes to the inverse gauge coupling strength. Perturbativity at the GUT scale set a bound on this quantity
\beqa
|\delta|\lesssim 24.3~.
\eeqa
with $24.3$ the value of $\al_{GUT}^{-1}$ without additional messengers. The SUSY scale is taken to be 2 TeV.

Proton decay experiments will also constrained the value of $\delta$. As the proton decay induced from the triplet Higgs depends on the scale of the triplets, we just take constraints from proton decay induced by heavy gauge bosons. The decay channel $p\ra \pi^0 e^+$ has the lifetime
\beqa
\tau(p\ra \pi^0 e^+)=\f{4 f_\pi^2 M_{X}^4}{\pi m_p\al_{GUT}^2(1+D+F)^2\al_N^2[A_{R;1}^2+(1+|V_{ud}|^2)^2A_{R;2}^2]}\[1+2\f{m_\pi^2}{m_p^2}\].
\eeqa
With updated experimental bounds from  Super-Kamiokande\cite{Kamiokande} $\tau>1.67\tm 10^{34}$ years, we can constraint the inputs
\beqa
\al_{GUT}\lesssim (5.27)^{-1}~,
\eeqa
by taking $f_\pi=131$ MeV, chiral Lagrangian factor $1+D+F=2.27$ with $D=0.80, F=0.47$\cite{proton}, the hadronic matrix element $\al_N=0.0112~{\rm GeV}^3$ (at renormalization scale $\mu=2 {\rm GeV}$) and $A_{R;1}=A_{R;2}\approx 5$. This value constrained $\delta$ to be
\beqa
|\delta|\lesssim 19~.
\eeqa

We should note that $A_{R;1}, A_{R;2}$, which represent the renormalization effects resulting from the anomalous dimensions of the operators, will also be amended by presence of vector-like messengers\cite{hisano}. They are defined as
\beqa
A_{R;1}=A_L A_{S;1}~,~~~A_{R;2}=A_L A_{S;2}~,
\eeqa
with $A_L,A_{S;i}$ the long and short distance factors, respectively. Here the long-distance contribution $A_L$ is taken to be $1.25$. The short distance factors will be changed into
\beqa
A_{S;i}&=&\(\prod\limits_{j}\f{\al_j(M_{SUSY})}{\al_j(M_Z)}\)^{\f{\gamma_{j;i}^{SM}}{b_j^{SM}}}
\(\prod\limits_{j}\f{\al_j(M_{mess})}{\al_j(M_{SUSY})}\)^{\f{\gamma_{j;i}^{MSSM}}{b_j^{MSSM}}}
\(\prod\limits_{j}\f{\al_j(M_{GUT})}{\al_j(M_{mess})}\)^{\f{\gamma_{j;i}^{MSSM}}{b_j^\pr}}~,\nn\\
&=&\(\prod\limits_{j}\f{\al_j(M_{SUSY})}{\al_j(M_Z)}\)^{\f{\gamma_{j;i}^{SM}}{b_j^{SM}}}
\(\prod\limits_{j}\f{\al_j(M_{GUT})}{\al_j(M_{SUSY})}\)^{\f{\gamma_{j;i}^{MSSM}}{b_j^{MSSM}}}
\(\prod\limits_{j}\f{\al_j(M_{GUT})}{\al_j(M_{mess})}\)^{\f{\gamma_{j;i}^{MSSM}}{b_j^\pr}-\f{\gamma_{j;i}^{MSSM}}{b_j^{MSSM}}}~,\nn\\
&\equiv& A_{S;i}^0\(\prod\limits_{j}\f{\al_j(M_{GUT})}{\al_j(M_{mess})}\)^{\f{\gamma_{j;i}^{MSSM}}{b_j^\pr}-\f{\gamma_{j;i}^{MSSM}}{b_j^{MSSM}}}.
\eeqa
in the case that one vector-like family of messengers at scale $M_{mess}$ are present. Results with multiple messenger thresholds can be trivially extended.
The relevant coefficients within the expressions are given\cite{ALAR} as
\beqa
\gamma_{j;1}^{SM}&=&(~2,~9/4,~11/20)~,~\gamma_{j;2}^{SM}=(~2,~9/4,~23/20)~,\nn\\
\gamma_{j;1}^{MSSM}&=&(~4/3,~3/2,~11/30)~,~\gamma_{j;2}^{SM}=(~4/3,~3/2,~23/30)~,
\eeqa
with $b_j$ the relevant gauge beta functions upon each threshold. The multiple factor for $A_{S;i}^0$ in the presence of messengers is given approximately by
 \beqa
F_1&=&\(\prod\limits_{j}\f{\al_j(M_{GUT})}{\al_j(M_{mess})}\)^{\f{\gamma_{j;i}^{MSSM}}{b_j^\pr}-\f{\gamma_{j;i}^{MSSM}}{b_j^{MSSM}}}\nn\\
 &\approx&\[1+\f{b_j^\pr}{2\pi}\al_j(M_{GUT})\ln\f{M_{GUT}}{M_{mess}}\]^{\f{\gamma_{j;i}^{MSSM}}{b_j^\pr}-\f{\gamma_{j;i}^{MSSM}}{b_j^{MSSM}}}~,\nn\\
  &\approx& 1-\f{\gamma_{j;i}^{MSSM}}{2\pi}\f{\Delta b_j^m}{b_j^{MSSM}}\al_j(M_{GUT})\ln\f{M_{GUT}}{M_{mess}}~,
 \eeqa
 in which we define $\Delta b_j^m=b_j^\pr-b_j^{MSSM}=n_1$. This multiple factor can be easily extended to include multiple messengers.
 For example, with additional messenger thresholds at $M_2$, the new multiple factor is given by
 \beqa
 F_2 &\approx& 1-\f{\gamma_{j;i}^{MSSM}}{2\pi}\f{n_2}{b_j^{MSSM}+n_1}\al_j(M_{GUT})\ln\f{M_{GUT}}{M_{2}}~,
 \eeqa
 with the total multiple factor
 \beqa
 F=\prod\limits_{k}F_k\approx 1-\f{\gamma_{j;i}^{MSSM}}{2\pi}\al_j(M_{GUT})\sum\limits_{k=1}^p\f{n_k}{b_j^{MSSM}+\sum\limits_{l=1}^{k-1} n_l}\ln\f{M_{GUT}}{M_{k}}~.
 \eeqa
As the coefficients $A_{R;1}, A_{R;2}$ depend on the messenger scales, the proton decay constraints will feed back into the constraints on $\delta$.
Detailed discussions on constraints for $\delta$ will be given in our subsequent studies.

\section{\label{conclusion}Conclusions}
Extraordinary gauge mediation extension of deflected AMSB scenarios can be interesting because it can accommodate together the deflection in the Kahler potential and the superpotential. We revisit the EGM scenario and derive the analytical expressions for soft SUSY breaking parameters in EGM and EGM extension of deflected AMSB scenarios with wavefunction renormalization approach, especially the case with vanishing gauge beta-function at an intermediate energy scale. We find in EGM that large hierarchy among the messenger thresholds may indicate non-negligible contributions to trilinear couplings at the lightest messenger threshold scale. The Landau pole and proton decay constraints are also discussed.


\appendix
\section{ Coefficients In the Wavefunction of MSSM Superfields}
From the anomalous dimension
\beqa
\f{d}{dt} \ln Z_{f}&=& \sum\limits_{l=g_3,g_2,g_1}2\tl{A}_l\f{d \ln g_l}{dt}+ \sum\limits_{l=y_t,y_b,y_\tau} 2B_l \f{d \ln y_{l}}{dt}~,\nn
\eeqa
in the basis of $(y_t^2,y_b^2,y_\tau^2,g_3^2,g_2^2,g_1^2)$, the coefficients $\tl{A}_l, B_l$ can be solved. Expressions of the coefficients had already been obtained in our previous paper\cite{analytic-mirage}. The coefficient $A_i$ are listed  in Table.\ref{Acoeff}.

\begin{table}[htbp]
\caption{The gauge field coefficients $A_i(i=1,2,3)$ within the wavefunction for MSSM superfields:}
\begin{center}
\begin{tabular}{|c|c|c|c|}
\hline
&$A_3(g_3)$&$A_2(g_2)$&$A_1(g_1)$\\
\hline
$Q_3$& $\f{128}{61}$&$\f{87}{61}$&-$\f{11}{61}$\\
\hline
$U_3$& $\f{144}{61}$&$-\f{108}{61}$&~$\f{144}{305}$\\
\hline
$D_3$& $\f{112}{61}$&$-\f{84}{61}$&~$\f{112}{305}$\\
\hline
$L_3$& $\f{80}{61}$&$~\f{123}{61}$&-$\f{103}{305}$\\
\hline
$E_3$& $\f{160}{61}$&$-\f{120}{61}$&~$\f{61}{32}$\\
\hline
$H_u$&~$-\f{272}{61}$&$~\f{21}{61}$&-$\f{89}{305}$\\
\hline
$H_d$&~$-\f{240}{61}$&$-\f{3}{61}$&-$\f{57}{305}$\\
\hline
$Q_2$& $\f{16}{3}$&~3&~$\f{1}{15}$\\
\hline
$U_2$&$\f{16}{3}$&~0&~$\f{16}{15}$\\
\hline
$D_2$&$\f{16}{3}$&~0&~$\f{4}{15}$\\
\hline
$L_2$&~0&~3&~$\f{3}{5}$\\
\hline
$E_2$&~0&~0&~$\f{12}{5}$\\
\hline
\end{tabular}
\end{center}
\label{Acoeff}
\end{table}

\begin{acknowledgments}
 This work was supported by the Natural Science Foundation of China under grant numbers 11675147,11775012.
\end{acknowledgments}

\end{document}